# The MACIV multiscale seismic experiments in the French Massif Central (2023-2027): deployment, data quality and availability

C. Aubert [1], G. Scheiblin [1], A. Paul [1], H. Pauchet [2], A. Mordret [1,3], V. Baudot [1], S. Chevrot [5], N. Cluzel [4], I. Douste-Bacqué [1], F. Grimaud [2], A. Jung [1], S. Mercier [1], P. Pawlowski [1], S. Roussel [1], T. Souriot [4], N.M. Shapiro [1], M. Sylvander [2], B. Vial [1], D. Wolyniec [1]

[1] Institut des Sciences de la Terre, Univ. Grenoble Alpes, Univ. Savoie Mont Blanc, CNRS, IRD, Univ. Gustave Eiffel, ISTerre, 38000 Grenoble, France
[2] Institut de Recherche en Astrophysique et Planétologie, UMR5277, Université Toulouse 3 – Paul Sabatier, CNRS, CNES, Toulouse, France
[3] Geological Survey of Denmark and Greenland, Geophysics and Sedimentary Basins Department, Copenhagen, Denmark
[4] Laboratoire Magmas et Volcans, CNRS, IRD, OPGC, Université Clermont Auvergne, 63000 Clermont-Ferrand, France
[5] Géosciences Environnement Toulouse, UMR 5563, Université Toulouse 3 – Paul Sabatier, CNRS, IRD, Toulouse, France

## Abstract

In the framework of the MACIV project, a consortium of French laboratories has deployed a temporary seismic network of 100 broadband stations in the French Massif Central (FMC) for 3-4 years (2023-2027). The project aims at imaging the crust and upper mantle of the FMC to better assess the sources of volcanism, and the impacts of the Variscan inheritance or the Cenozoic rift system on volcanic systems. A large-scale array of 35 broadband stations covers the entire FMC and complements the permanent networks to reach a homogeneous coverage with ~35 km spacing. This network, with XP code, is the French contribution to AdriaArray. The XP array is complemented with 3 quasi-linear north-south, east-west and northwest-southeast profiles with inter-station spacing of 5-20 km, making up the XF network of 65 stations. The profiles cross volcanic areas and the main Variscan structures. We describe the experimental setup designed to optimize the performance/cost ratio and minimize the number of field visits, the deployment, the state-of-health monitoring, the data management and the data quality control strategies, outcomes of our 15-years' experience with major temporary seismic experiments in France and neighboring countries, including AlpArray. We also show some preliminary results including hypocenter locations and receiver function analysis. The 2 broadband arrays will be supplemented in 2025 by a month-long deployment of 3 large-N dense arrays of 625 3-C short-period nodes. These dense arrays will complete our multi-scale seismic experiment and illuminate active faults and possible plumbing systems of the youngest volcanoes.

# 1. Introduction

The MACIV project aims at probing the crustal and upper mantle structure with passive seismic imaging methods and improving seismic monitoring in the French Massif Central (FMC) using multi-scale temporary seismic experiments. To improve the coverage of the FMC by seismic networks, we have installed 100 temporary stations in 2023 and 2024. The entire process of defining the experimental set-up, strategies of site selection, deployment, health monitoring, data management and data quality control is based on a 15 years' experience of large-scale temporary seismic experiments by the ISTerre (Univ. Grenoble Alpes) and IRAP-GET (Univ. Toulouse) teams involved in the MACIV project.

The two teams started collaborating for the PYROPE seismic experiments in the Pyrénées (2010-2014), in a project led by the Toulouse group (e.g. Chevrot et al., 2014; Dataset: Chevrot et al., 2017). PYROPE was our first deployment of a large-scale broadband temporary network in France, without established procedure for site selection, installation, quality control, etc. The Grenoble group progressively improved procedures in the CIFALPS, CIFALPS2, and AlpArray-FR projects in the Western Alps between 2012 and 2020 (e.g. Zhao et al., 2015; Nouibat et al., 2022; Paul et al., 2022; Datasets: AlpArray Seismic Network, 2015; Zhao et al., 2016, 2018). Based on the CIFALPS experience of setting up and sharing installation procedures with the involved INGV teams, the ISTerre group took an active part in the definition of the shared technical strategy for the AlpArray temporary seismic array (Brisbourne et al., 2013). The use of GSM communication for real-time data transmission and state-of-health monitoring started for AlpArray-FR and it was generalized for CIFALPS2. The large number of instruments, the extensive spatial spread, the long duration of experiments (e.g., AlpArray-FR: 80 temporary stations over Eastern France for more than 3 years), and the sharing of installation work with several other laboratories in France and Italy have led us to develop and standardize our procedures for site search, installation, state-of-health monitoring, data transmission, etc.

We have further improved these procedures for the new MACIV experiment, with its 100 broadband stations spread across the FMC, for instance by developing new study instrumentation boxes. These boxes make installation easier and faster for small, non-expert teams, and their robustness prevents problems requiring on-site maintenance. These new boxes, combined with reliable seismic instruments and GSM communication for efficient state-of-health monitoring, help to improve network performance at a reasonable cost by minimizing the number of maintenance visits, which saves time, money and $CO_2$ emissions. We have never described these procedures in a publication, apart from a summary for CIFALPS2 in the supplementary information file of Paul et al. (2022). With MACIV, the culmination of 15 years of constant improvements, we are taking the opportunity of this special issue to fill that gap and share our experience.

# 2. The MACIV project and temporary seismic experiments: presentation

## 2.1 The MACIV project: aims and scopes

MACIV stands for "Multi-scale seismic imaging of Massif Central focussing on recent intraplate volcanism". It is a collaborative research project involving ISTerre (Univ. Grenoble Alpes), GET & IRAP (Univ. Toulouse) and LMV (Univ. Clermont Auvergne). MACIV is funded by the French national research funding agency ANR (Agence nationale de la recherche).

The goal of MACIV is to obtain new information about the crustal and lithospheric structures and ongoing volcanic processes beneath the FMC. The FMC occupies a strategic position in the French geological and geographical landscape, having recorded a varied geological history from the end of the Proterozoic to the present, i.e., nearly 600 Ma. MACIV specifically aims to improve our understanding of the impact of the lithological-structural inheritance from the Variscan history and the role of present-day geodynamics on the FMC Cenozoic intraplate volcanism.

The European volcanic provinces, and the FMC specifically, exhibit clear links with the Cenozoic rift systems and the European Variscan belt, but the specific role of each structure on their origin, dynamics, and morphology is still poorly documented. The current activity of the Eifel volcanic region (Germany; Hensch et al., 2019), a system similar to the FMC, reminds us of the severe natural hazards that remain difficult to assess in these dormant volcanic regions.

The backbone of the MACIV temporary array is the westernmost part of AdriaArray, a multi-national effort to cover the Adriatic Plate and its active margins by a dense array of seismic stations to understand the causes of active tectonics and volcanic fields (Kolínský, Meier, et al., 2025).

## 2.2 The MACIV multiscale seismic experiments

Figure 1 shows a map of the MACIV temporary arrays and permanent seismic networks in the FMC (FR: Epos-France Seismology, 1995; RD: RESIF, 2018; G: IPGP & EOST, 1982; RA: RESIF, 1995). The temporary network includes 3 sub-arrays of different scales to efficiently probe the scales and depths of the FMC lithospheric structures and volcanic systems.

- The MACIV backbone array covers the whole FMC and supplements the French permanent broadband network of Epos-France with 35 broadband temporary stations (blue open triangles in Fig. 1b-c). Sites were selected to fill gaps in the permanent network and reach a uniform spatial coverage with an average station spacing of 35 km. Stations have been installed between March 2023 and January 2024, and they will operate until late 2027. This backbone aims at imaging the crust and upper mantle structures at deca-kilometric resolution. It is the French contribution to AdriaArray. Data is publicly available in real time from the Epos-France EIDA node.
- Three transects of intermediate-band (IB) stations oriented north-south, east-west and northwest-southeast complement the backbone to probe crustal structures at kilometric to deca-kilometric resolution (orange diamonds in Fig. 1b-c). Station spacing decreases from 20 km at the ends of the transects to 5 km in the central part. Temporary IB stations have been deployed in 2024 and will operate until late 2027. (1) The north-south transect is 420-km long and includes 41 stations in total, with 30 IB stations and 11 backbone and permanent stations; it crosscuts the main Variscan sutures and crustal blocks, and goes along the main volcanic centers of the Chaine des Puys, Cézallier-Mont-Dore and Cantal. (2) The east-west transect is 370-km long and includes 31 stations, with 26 IB stations; It crosses the main Variscan nappes and is orthogonal to the Sillon Houiller, the Chaine des Puys and the Limagne grabens. (3) The so-called diagonal transect is 270 km-long and oriented northwest-southeast; It includes 23 stations in total, with 9 IB stations; It crosses the volcanic centers of Velay, Devès and Bas-Vivarais. Data is embargoed for 2 years after installation, until July 2026.
- Three dense nodal arrays of 625 short-period 3-component instruments will be deployed for one month in the autumn of 2025 in the Chaine des Puys and Cézallier-Mont-Dore areas. They will densify the backbone and transects in the regions that experienced the latest volcanic activity. Their data will be used to image the shallow subsurface at kilometric scale. We will not describe this part of the MACIV arrays in this paper because it is still under construction at the time of writing.

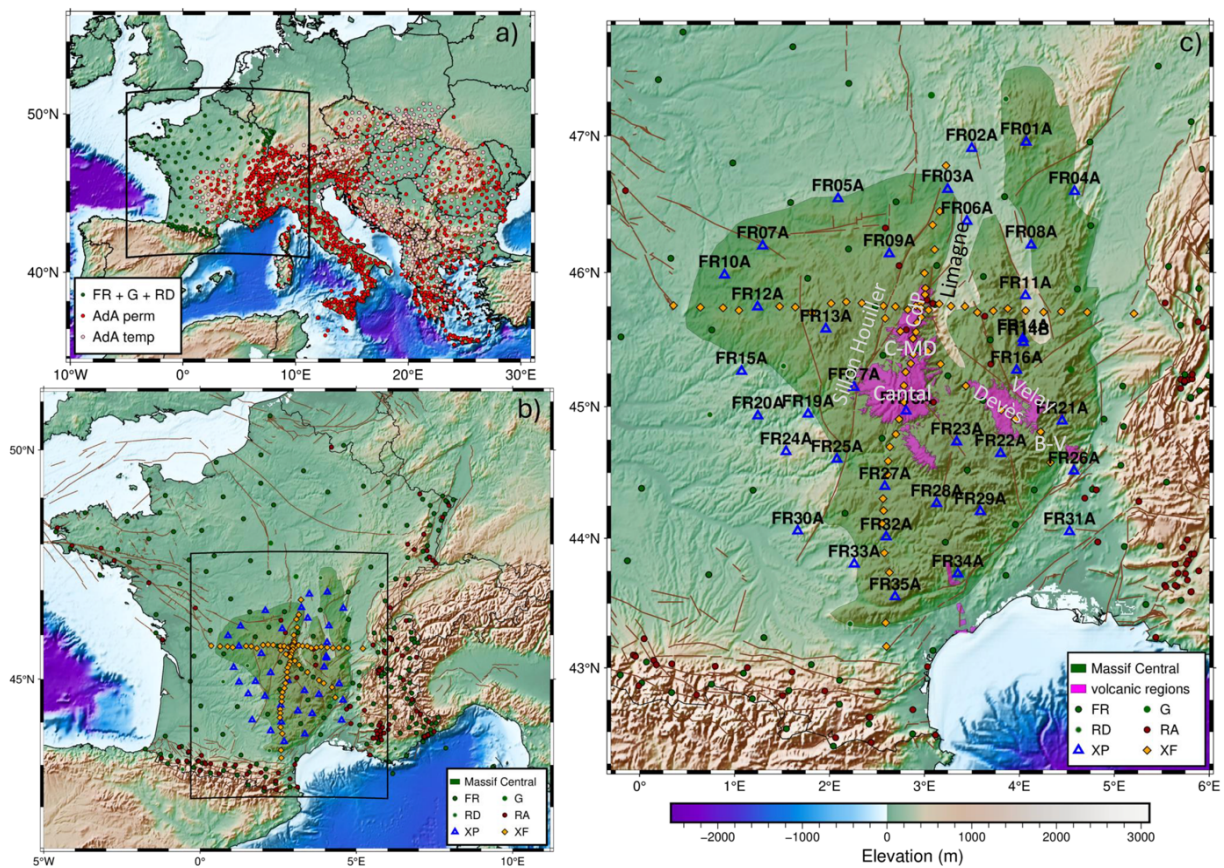

**Figure 1.** Maps of MACIV broadband seismic stations. (a) At European scale, map of AdriaArray (AdA) permanent (red circles) and temporary (pink circles) stations, and permanent stations in France (green circles). (b) Zoom-in on France showing permanent stations (networks FR, RD, G, RA), and MACIV temporary stations (blue triangles: XP stations, orange diamonds: XF stations). The green-filled area is the FMC Variscan massif. (c) Zoom-in on the FMC showing locations of permanent and temporary stations, major faults, and volcanic regions (pink-filled areas). B-V: Bas Vivarais; CdP: Chaine des Puys; C-MD: Cézallier-Mont Dore.

# 3. The MACIV backbone and profiles: infrastructural setting, site selection, installation, data transmission

## 3.1 Station and vault design

The 100 stations part of the MACIV project use instruments of the French pool of mobile seismic instruments SisMob (https://sismob.epos-france.fr/).

For such a large number of temporary stations in multi-scale arrays, station infrastructures were standardized as much as possible. The instrument boxes were designed specifically for the MACIV project, with the aim of optimizing installation time, facilitating implementation, and improving the robustness of the station. They are hermetically sealed and integrate all components required for data acquisition, real-time transmission and power supply. They were assembled in the laboratory prior to on-site deployment. Their manufacturing cost (580 € including supplies for sealing and elements for safe and reliable electrical wiring, plus half a day of manpower) is low compared to the number of days saved in the field, both for installation and maintenance. The instrument boxes were designed according to station type, broadband (BB) for the backbone or intermediate-band (IB) for the transects, and according to site type, outdoor or indoor.

A BB station consists of: (1) a broadband sensor with low corner frequency 120s (Trillium 120QA, Horizon or Compact 120 series or STS-2); (2) a Centaur 3-channels and 24 bits digitizer (CTR4 or CTR3 series) with a dynamic range up to 142 dB at 100 sps sampling rate; (3) an external GNSS antenna for clock synchronization; (4) a NB1600 LTE Mobile router for real-time data transmission equipped with an external antenna 4G-LTE (omnidirectional or directional, depending on site exposure and reception levels of on-site mobile operators).

For BB stations, sites were mainly inside buildings with access to mains power. This ensured safety and optimum installation of BB sensors, which are not hermetically sealed. To ensure thermal insulation of the sensor, we used a concrete slab of 40x40 $cm^2$, with a polyvinyl chloride tube with a screw-on lid inserted into the slab. Absorbent cotton panels were inserted around the sensor to limit air convection. Insulation was completed with an external multi-layer aluminum sheet (Fig. 2a to f).

To ensure continuity of data acquisition for 10 days in case of power cut, we installed two sealed lead acid batteries, 12V-65Ah each. Finally, a battery protect module was integrated in the electrical circuit, which disconnects the batteries from non-essential loads before they are completely discharged. All the instrumentation and electrical components were integrated and wired in the instrument case, ready for on-site deployment (Fig. 2g, h). In sites where mains power supply is not available (2 of 35), the station is powered by a solar panel of 115W peak power, 18V and two batteries of 65Ah (Fig. 3).

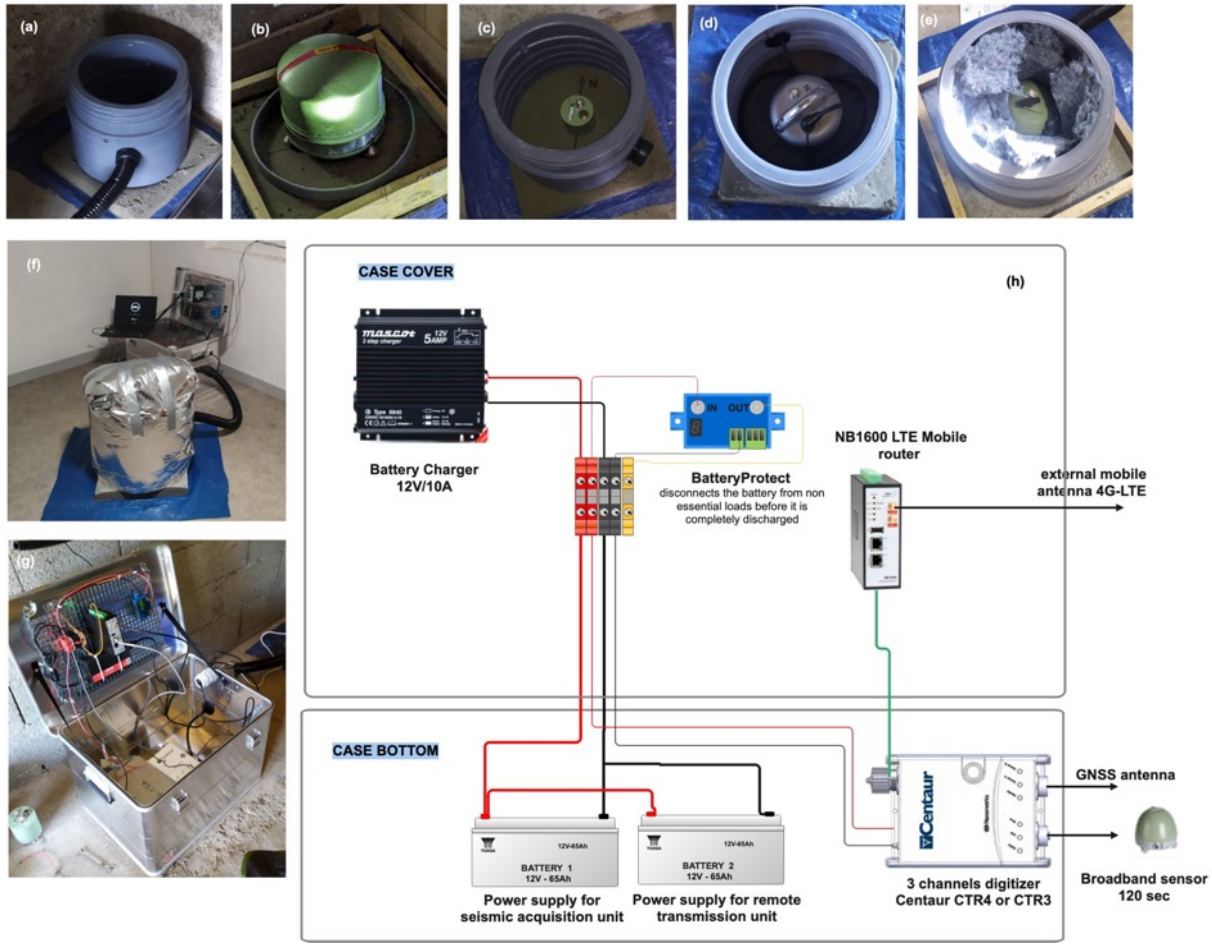

**Figure 2.** Components of a broadband seismological station installed on the MACIV backbone network (XP). (a) Sensor vault; (b) STS-2 sensor; (c) Trillium Compact 120 series sensor; (d) Trillium Horizon sensor; (e) Trillium 120 QA sensor in its vault, surrounded with absorbent cotton panels; (f) Sensor vault fully thermally insulated; (g) Instrument box; (h) wiring diagram.

An IB station has the same components as a BB station, with the BB velocimeter replaced by an intermediate-band velocimeter with low corner frequency 20s (Trillium Compact Posthole 20 sec - TC20-PH2).

The IB stations of the transects were mainly installed outdoor, close to homes or properties for safety reasons (45 stations out of 65). Outdoor sites were preferred because sensors are designed for posthole installation (Fig. 3a, b). Other stations were installed inside buildings, following the same procedure as for BB stations (Fig. 2).

For outdoor installation, the sensor is buried in a hole at least 50 cm deep, drilled with a 150 mm diameter auger, slightly larger than the 97 mm diameter of the sensors (Fig. 3c, d). To stabilize the sensor and facilitate levelling, sand is poured at the bottom of the hole. The sensor is then oriented to the north with a compass. Levelling is performed using a spirit level placed on top of the sensor and the digital spirit level accessible via the sensor web page from the digitizer. Sand is poured all around the sensor to aid levelling and stabilize its position (Fig. 3e). Finally, the hole is filled with soil.

Stations are powered by a solar panel of 115W peak power, 18V and three batteries 65 Ah each. The solar panel is mounted on a tilting structure with adjustable feet (Fig. 3f). The solar controller (SunSaver Duo-25) allows to dissociate battery packs and to power the modules independently: the data acquisition part (digitizer and sensor with 1.5W total consumption) is powered by a 130Ah battery pack, and the remote transmission part (router with 4W consumption) is powered by a 65Ah battery pack. This setting increases the autonomy of the acquisition part (to around 40 days) in case of bad weather or theft of panels, while the teletransmission part only has an 8-day autonomy. A battery protect module disconnects the batteries from non-essential loads before they are completely discharged. All the instrumentation and electrical components are integrated and wired in the instrument case, ready for on-site deployment (Fig. 3g, h).

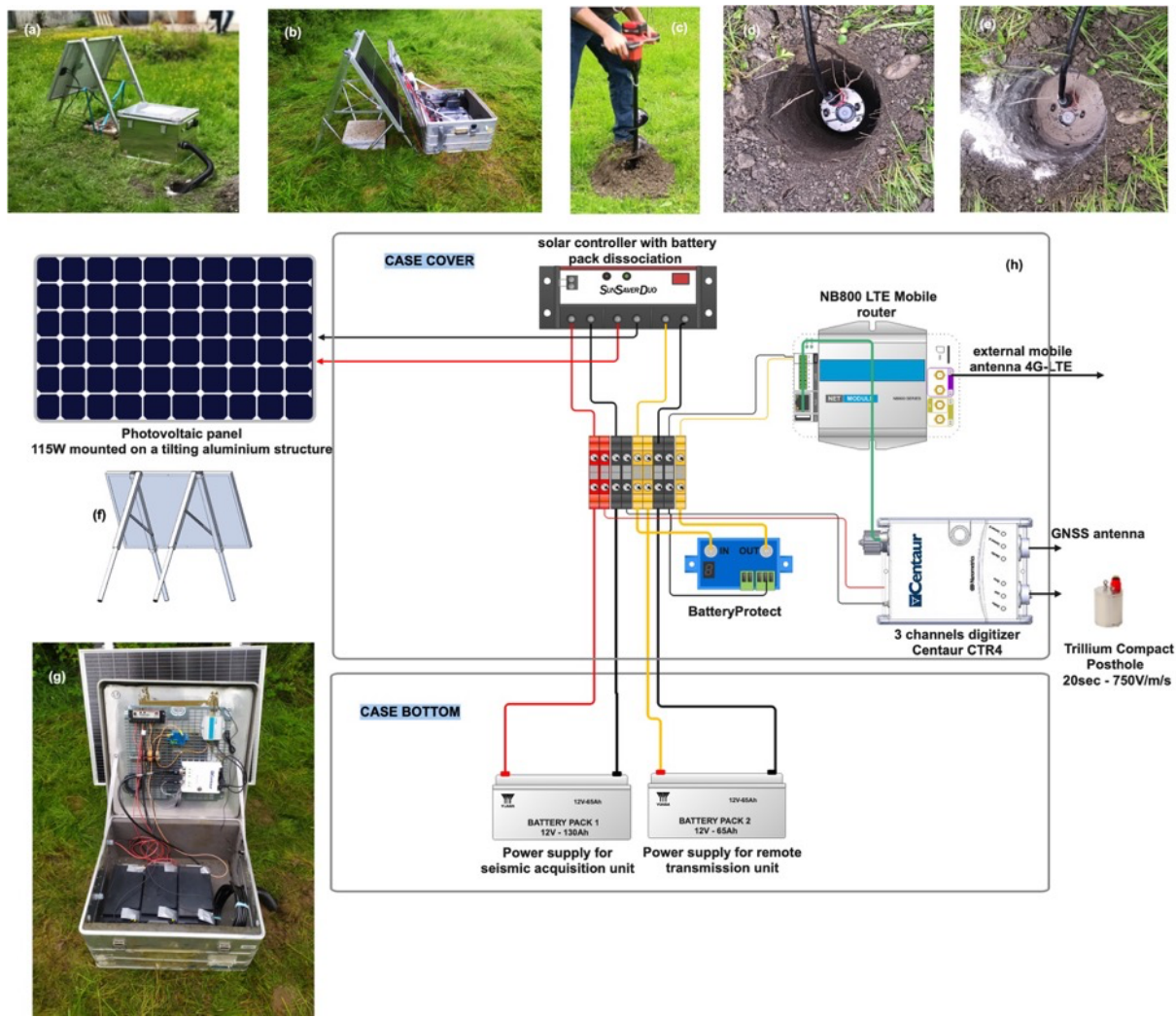

**Figure 3.** Components of an intermediate-band seismological station installed on the transects (XF). (a, b) Typical outdoor station installation; (c) Auger digging sensor hole; (d, e) Trillium Compact Posthole 20 s - TC20-PH2 sensor positioned in its hole; (f) Tilting structure used for solar panel; (g) Instrument box; (h) Wiring diagram.

### 3.2 Site scouting

Building on previous experience, we defined a robust procedure to search for suitable sites for temporary stations. We first positioned theoretical station locations on a map using tools such as GoogleEarth or the French geographic portal Geoportail (https://www.geoportail.gouv.fr/), with a circle (or ellipse) defining the maximum possible deviation from the ideal location. Geology was roughly checked to identify soil types. Municipalities included in the circle were then listed for each station location. A set of possible sites was selected for each location, far enough from sources of major anthropogenic or natural noise (highways, railroads, intensive industrial or agricultural facilities, streams, forests, etc...).

A leaflet designed for the general public (town hall, private owners) that explains the scientific project, describes a seismological station with photos and indicates installation constraints, was sent by email to all town halls identified in the search areas. Then, local authorities (mayors) were systematically contacted by phone. Discussion was generally easy thanks to the leaflet. Town halls often suggested sites belonging to the municipality, such as cellars, storage areas, or unused churches or chapels. When no municipal site was suitable, mayors put us in touch with private owners or associations. In-depth phone discussions with municipalities or private landowners helped us to identify potential noise sources that could alter data quality.

The GSM coverage was also discussed with mayors and landowners, and checked on the website of the national agency of frequencies (Agence Nationale des FRéquences, https://www.cartoradio.fr), which maps radioelectric sites and measurements. For outdoor sites, we checked the sunlight potential of solar panels. Finally, site access and safety were checked, again by phone.

When possible, proposed locations were visited during dedicated surveys prior to installation. Such surveys have been carried for the profiles, when 2-3 stations could be scouted per day. When site survey could not be carried out beforehand, it was performed on the same day as the installation. In such cases, we had selected several sites beforehand, which were visited on the same day with owners or local authorities.

### 3.3 Installation

The BB backbone and the IB profiles were installed jointly by ISTerre (74 stations) and by IRAP/GET (26 stations), with technical help from LMV (Clermont-Ferrand). Each team (ISTerre, IRAP/GET) carried out site scouting and installation independently. The 35 stations of the backbone (XP network code) were installed between Feb. 28, 2023 and Jan. 24, 2024, with 74% of the stations installed before July 2023. They are listed in Table 1.

We carried out site survey and station installation on the same day to minimize travel, because sites were often more than 50 km apart, and several hundred km from Grenoble or Toulouse. We re-used six sites of the Alparray-FR project (2015-2020) (Table 1).

Installing an indoor broadband station required one day in average for 2 people, including travel, final site search, and installation. We used a cellular signal analyzer (SNYPER-LTE+ spectrum) dedicated to surveying the 4G/LTE European network to determine which mobile operators were present on site, as well as signal reception levels and data rates. This tool is crucial to define the type of mobile antenna to be installed, omni-directional when mobile coverage is good, or directional when it is weak. In the last case, the analyzer could be used to optimize antenna positioning. Detailed examples of 2 BB stations are given in Figure 4a, b.

| Station Name | Lat | Lon | Elev (m) | Site Name | StartTime (yyyy-mm-ddThh:mm:ss) | EndTime (yyyy-mm-ddThh:mm:ss) | HousingClass | Sensor burried | sensor sits on | sensor type | digitizer type |
|---|---|---|---|---|---|---|---|---|---|---|---|
| FR01A | 46.9556 | 4.0720 | 457 | Saint Prix - 71 | 2023-06-06T08:00:00 | | Building | no | cement | STS2 | CTR4-3S |
| FR02A | 46.9102 | 3.4984 | 271 | Thianges | 2023-06-05T13:15:00 | | Building | no | cement | STS2 | CTR4-3S |
| FR03A | 46.611798 | 3.245900 | 239 | Montilly - 03 | 2023-06-06T14:00:00 | | Building | no | cement | STS2 | CTR4-3S |
| FR04A | 46.5963 | 4.5818 | 300 | Burzy - 71 | 2023-02-28T14:00:00 | 2024-02-14T10:00:00 | Building | no | cement | STS2 | CTR3-3S |
| FR04A | 46.5963 | 4.5818 | 300 | Burzy - 71 | 2024-02-14T11:00:00 | | Building | no | cement | TC 120 | CTR3-3S |
| FR05A | 46.543582 | 2.090276 | 252 | La Motte Feuilly - 36 | 2023-06-19T17:00:00 | | Building | no | cement | T120QA | CTR4-3S |
| FR06A | 46.3781 | 3.4474 | 140 | Saint Gerand De Vaux - 03 | 2023-06-06T09:00:00 | | Building | no | cement | STS2 | CTR4-3S |
| FR07A | 46.196386 | 1.296828 | 841 | Saint Sornin Leulac - 87 | 2023-10-04T10:00:00 | | Building | no | cement | T120QA | CTR4-3S |
| FR08A | 46.2034 | 4.1241 | 410 | Fleury la montagne - 71 | 2023-06-07T10:30:00 | | Building | no | cement | STS2 | CTR4-3S |
| FR09A | 46.13818 | 2.62999 | 550 | Saint Fargeol - 03 | 2023-06-20T16:00:00 | | Building | no | cement | STS2 | CTR4-3S |
| FR10A | 45.9823 | 0.895067 | 289 | Montrollet - 16 | 2023-06-20T10:00:00 | | Building | no | cement | T120QA | CTR4-3S |
| FR11A | 45.829039 | 4.064836 | 184 | Pommiers-en-Forez-42 | 2024-01-24T15:00:00 | | Building | no | cement | TH120-2 | CTR4-3S |
| FR12A | 45.745156 | 1.242994 | 289 | Solignac - 87 | 2023-06-20T15:00:00 | | Building | no | cement | T120QA | CTR4-3S |
| FR13A | 45.581207 | 1.961757 | 780 | Pérols sur Vézère - 19 | 2023-10-05T08:00:00 | | Building | no | cement | TH120-2 | CTR4-3S |
| FR14A | 45.510282 | 4.038651 | 740 | Soleymieux - 42 | 2023-12-06T14:00:00 | 2024-01-24T16:00:00 | Building | no | cement | STS2 | CTR4-3S |
| FR14B | 45.481489 | 4.041772 | 850 | Marols - 42 | 2024-01-25T11:00:00 | | Building | no | cement | STS2 | CTR4-3S |
| FR15A | 45.265319 | 1.074519 | 184 | Tourtoirac – 24 | 2023-07-25T15:00:00 | | Building | no | tiles | T120QA | CTR4-3S |
| FR16A | 45.273228 | 3.970473 | 841 | Saint André de Chalencon - 43 | 2023-10-05T11:30:00 | | Building | no | cement | STS2 | CTR4-3S |
| FR17A | 45.1452 | 2.2587 | 670 | Barriac les bosquets - 15 | 2023-06-21T11:00:00 | | Building | no | cement | T120QA | CTR3-3S |
| FR18A | 44.967202 | 2.807042 | 840 | Brezons - 15 | 2023-04-18T12:00:00 | | Building | no | cement | TC 120 | CTR4-3S |
| FR19A | 44.944067 | 1.774477 | 201 | Bilhac – 19 | 2023-07-24T15:00:00 | | Building | no | tiles | T120QA | CTR3-3S |
| FR20A | 44.930882 | 1.246428 | 184 | Proissans – 24 | 2023-07-25T12:00:00 | | Building | no | tiles | T120QA | CTR3-3S |
| FR21A | 44.892 | 4.4516 | 750 | Saint Michel d'aurence - 07 | 2023-04-06T10:00:00 | | Building | no | cement | STS2 | CTR4-3S |
| FR22A | 44.646349 | 3.801775 | 1168 | Cheylard-l'évêque – 48 | 2023-08-30T16:00:00 | | Building | no | tiles | T120QA | CTR3-3S |
| FR23A | 44.7343 | 3.3398 | 935 | Rimeize - 48 | 2023-04-20T10:30:00 | | Building | no | cement | T120QA | CTR4-3S |
| FR24A | 44.660405 | 1.542714 | 355 | Coeur-de-Causse – 46 | 2023-07-10T17:00:00 | | Building | no | tiles | T120QA | CTR3-3S |
| FR25A | 44.60301 | 2.079238 | 240 | Lunan – 46 | 2023-07-11T12:00:00 | | Building | no | tiles | T120QA | CTR3-3S |
| FR26A | 44.514045 | 4.574869 | 208 | Valvignères - 07 | 2023-04-05T10:00:00 | 2023-09-20T09:00:00 | Building | no | cement | STS2 | CTR4-3S |
| FR26A | 44.514045 | 4.574869 | 208 | Valvignères - 07 | 2023-09-20T11:00:00 | | Building | no | cement | TC 120 | CTR4-3S |
| FR27A | 44.396261 | 2.583255 | 600 | Onet-le-Château - 12 | 2023-07-11T17:00:00 | | Building | no | tiles | T120QA | CTR3-3S |
| FR28A | 44.266222 | 3.129464 | 875 | Massegros-Causses-Gorges – 48 | 2023-07-12T17:00:00 | | Building | no | tiles | T120QA | CTR3-3S |
| FR29A | 44.205912 | 3.587366 | 739 | Rousses – 48 | 2023-08-31T11:00:00 | | Building | no | tiles | T120QA | CTR3-3S |
| FR30A | 44.0562 | 1.6665 | 240 | Bruniquel -82 | 2023-06-14T10:00:00 | | Building | no | tiles | T120QA | CTR3-3S |
| FR31A | 44.0515 | 4.5280 | 259 | La Capelle et Masmolène - 30 | 2023-04-04T11:00:00 | | Free field | no | cement | TC 120 | CTR4-3S |
| FR32A | 44.011403 | 2.597037 | 415 | Connac – 12 | 2023-09-01T11:00:00 | | Building | no | tiles | T120QA | CTR3-3S |
| FR33A | 43.803309 | 2.261274 | 340 | Terre-de-Brancalié – 81 | 2023-10-23T11:00:00 | | Building | no | tiles | T120QA | CTR3-3S |
| FR34A | 43.72759 | 3.351946 | 404 | Soumont – 34 | 2023-07-13T12:00:00 | | Free field | no | tiles | T120QA | CTR3-3S |
| FR35A | 43.551150 | 2.691646 | 946 | Le Soulié – 34 | 2023-06-28T12:00:00 | | Building | no | tiles | T120QA | CTR3-3S |

**Table 1.** List of MACIV backbone stations (network code XP) installed between February 2023 and January 2024, with station name, coordinates, start and end time, type of housing, sensor type (TC 120: Trillium Compact 120 s, T120QA: Trillium 120QA, TH120-2: Trillium Horizon), digitizer type (CTR3-3S: Centaur3 series 3-channels, CTR4-3S: Centaur4 series 3-channels). In grey, stations operated by ISTerre, Grenoble. In blue, stations operated by IRAP, Toulouse. Sites already used in the AlpArray-FR project are shown in bold.

We applied a different deployment strategy for the 65 stations spread over the 3 profiles. As the inter-station distance is shorter (5 km for the inner part and 20 km for the outer parts of the profiles), several teams worked for several consecutive days to install as many stations as possible in the shortest possible time. Installation times were optimized because sites had been scouted beforehand, and 70% of the stations were installed outdoor and powered by solar panels. A training had been organized for all field workers and a standard procedure for the installation of IB stations was drawn up so that each team could be autonomous in the field. The 53 IB stations operated by the ISTerre were installed in 2.5 weeks by 3 or 4 teams of 2/3 persons between June 2024 and October 2024. The 12 IB stations operated by IRAP are currently being installed, by a single team of 2 people for 7 days. Completion of the installations is scheduled by the end of January 2025. Detailed examples of 2 IB stations are given in Figure 4c, d.

Installing an outdoor IB station takes 3-4 hours thanks to the burial of the posthole sensor directly into the ground, and to the use of an accumulator-mounted auger to dig the hole. The sensor cable is passed through a ringed sheath, connected to the instrument box by a cable gland to complete instrument protection. The solar panel is installed on its free-standing structure secured to the ground by 4 galvanized steel anchor bolts and 25 kg cement weights.

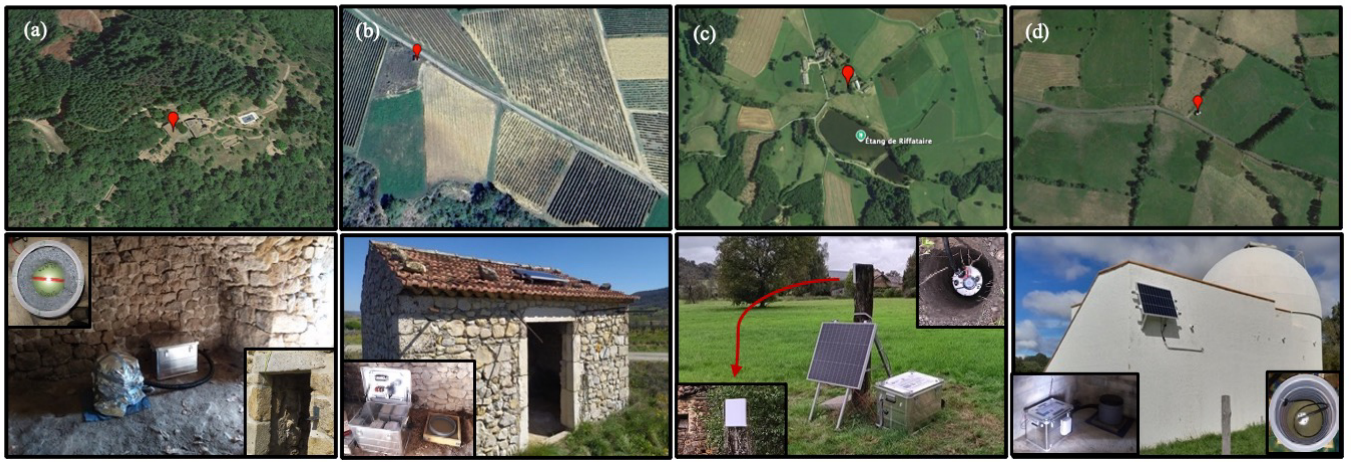

**Figure 4.** Examples of BB and IB stations installed as part of the MACIV project. (a) Station XP.FR21A is installed in the half-buried basement of a holiday home, with a dirt floor on the rock. (b) Station XP.FR26A is installed in a barn at the edge of an uncultivated field and is powered with a solar panel installed on the roof. (c) Station XF.ME06A is installed in a private field. (d) Station XF.MN19A is installed in a community astronomical observatory.

### 3.4 Data transmission and distribution

All data from the backbone network (network code XP; dataset: Paul et al., 2023) and the 3 profiles (network code XF; dataset: Paul et al., 2024) are transmitted in real time through the mobile communication system. The digitizer and LTE mobile router are connected at each station to form a local network. The modem-router, via the mobile operator, obtains a dynamic IP address that evolves over time. To guarantee continuous data transmission, a dynamic DNS service is used to associate the dynamic IP address with a single and unique domain name. Each element of the local network is also remotely accessible via a port forwarding service. The whole system is secured by a local firewall (Fig. 5a).

Data streams are transmitted to a Virtual Machine (VM) dedicated to real-time data, which is monitored by the Geophysical Instrumentation Service (SIG) of ISTerre and hosted in the IT infrastructure of Grenoble observatory (OSUG). Real-time seedlink streams (network protocol for near-real-time data collection and distribution in miniSEED format) are collected using a *slarchive* client installed on the real-time server (Fig. 5b). Data is then archived in a SDS structure (Data Structure for archiving miniSEED waveform data) on SUMMER, the shared storage platform of Grenoble Alpes University (UGA). This step is performed using the *slarchive* client integrated into the SeisComp software installed on the SC3 Virtual Machine (Fig. 5c). To configure the SeisComp software to monitor these incoming data flows, we build a temporary stationXML file for each station using Yasmine (Yet Another Station Metadata INformation Editor), a Python web application that creates seismological station metadata information in FDSN stationXML format, and the *yasmine-cli* command-line script (Saurel et al, 2020). At this stage, a data completion step is carried out on a daily basis using an in-house program. The last step is data transfer to the Epos-France SI-S data center for archiving and distribution, which is also located in OSUG and UGA.

The distribution rules of the MACIV datasets differ according to network code. The XP dataset (BB backbone) is publicly available, while the XF dataset (IB profiles) is embargoed until July 2026. Real-time and completed data of the XP network are transferred to the Epos-France SI-S data center, while only completed data is transferred for XF (Fig. 5d). Apart from the real-time data flow of XP, MACIV data is synchronized twice a day on the Epos-France datacenter (Fig. 4e). As described in Péquegnat et al. (2021), data integration at the Epos-France SI-S data center is carried out in two stages. (1) Data is collected by the SisMob A-node, which is responsible for data integration, validation and metadata construction in FDSN stationXML format; (2) it is transmitted to the "B-node", which is responsible for data and metadata archiving and distribution using FDSN Webservices (Fig. 5f-g).

The SC3 VM is also configured for routine location of Alpine earthquakes (Fig. 5c). A second instance dedicated to earthquake location in the FMC has been installed on a dedicated virtual machine (SCMCF). This instance benefits from the automatic location procedure performed on the SC3 VM. Events identified as "out of zone" on the VM dedicated to the Alps are automatically processed on the VM dedicated to the FMC, and manually checked (Fig. 5h).

Finally, real-time data streams are also transmitted to the data server hosting SYNAPSE, the supervision tool developed and monitored by OCA (Côte d'Azur Observatory in Nice) for permanent and temporary seismological stations of Epos-France (Fig. 5i).

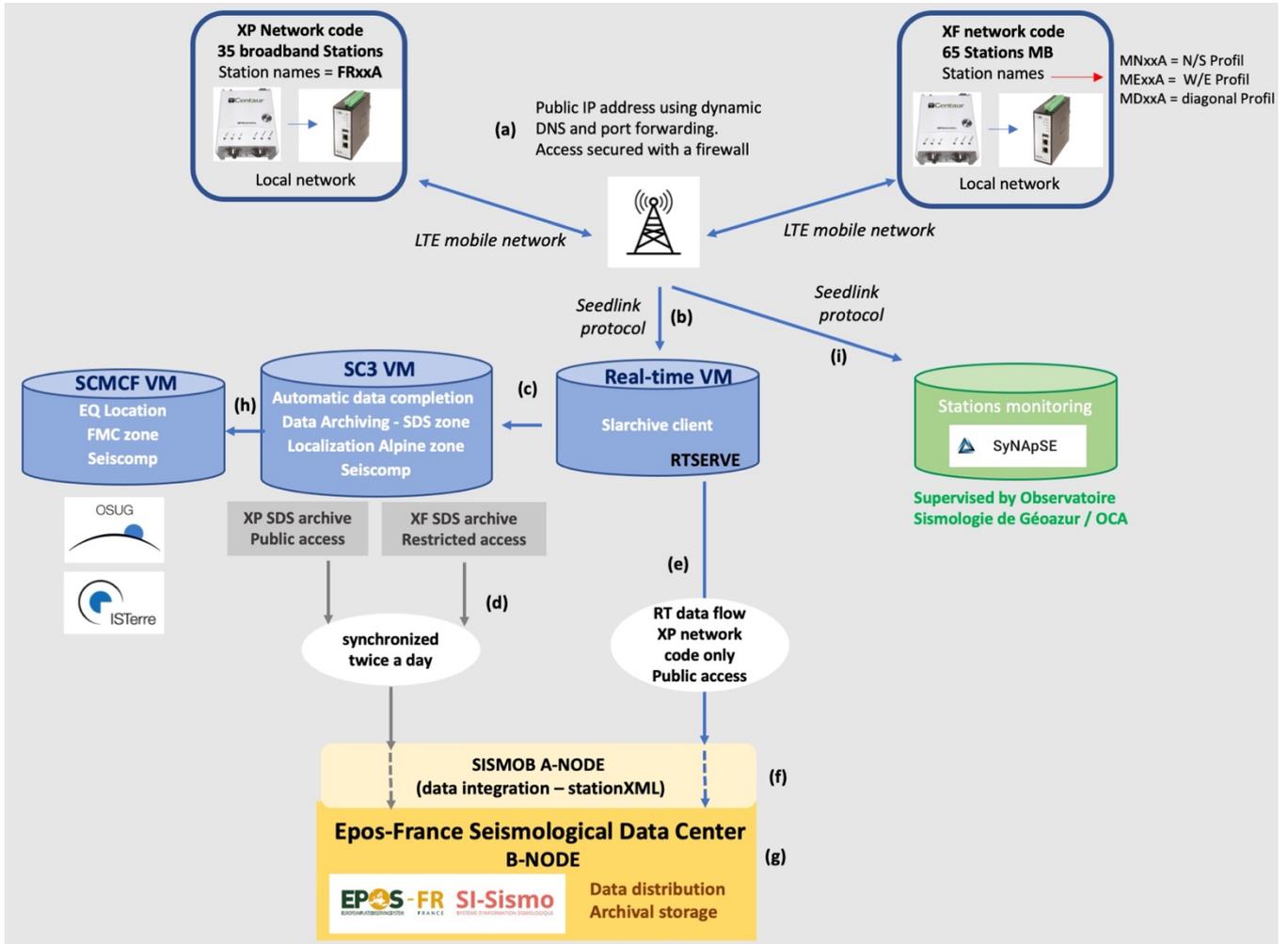

**Figure 5.** Diagram illustrating the seismological data flows of the MACIV project. (a) Protocol for connecting the seismological station to the operator's mobile network. (b) Transfer of data flows to the VM dedicated to real time. (c) SDS architecture data archiving on SUMMER (UGA) and automatic data completion. (d) Synchronization of completed data to Epos-France SI-S datacenter. (e) Real-time data transfer to Epos-France SI-S datacenter. (f) SISMOB A-Node. (g) Epos-France Seismological datacenter B-node. (h) VM for earthquake (EQ) location in FMC. (i) Real-time data transfer to SYNAPSE monitoring tool.

### 3.5 Daily monitoring and Data availability

The operation of XP and XF stations (backbone and profiles) is remotely supervised on a daily basis using the SYNAPSE supervision tool. This tool was developed by OCA (Côte d'Azur Observatory, Nice) in 2016 and designed to monitor the operation of permanent networks (https://www.epos-france.fr/en/scientific-promotion/software-and-tools/). A simple color code is used to quickly identify the station status. All supervised parameters are displayed in graphical form and can be easily viewed by accessing the individual station page. Alert thresholds can be configured at operators' request. The main elements monitored are: real-time or near-real-time data flow to the Epos-France SI-S data center, modem status (ping time, packet lots, cell ID and operator name, signal strength), seismic digitizer status (battery voltage, timing quality, number of satellites, system power consumption, temperature) and sensor mass position status. With this tool and the daily checks by project staff, failures are easily detected and quickly fixed by triggering maintenance operations either remotely when possible, or on-site when necessary (Fig. 6).

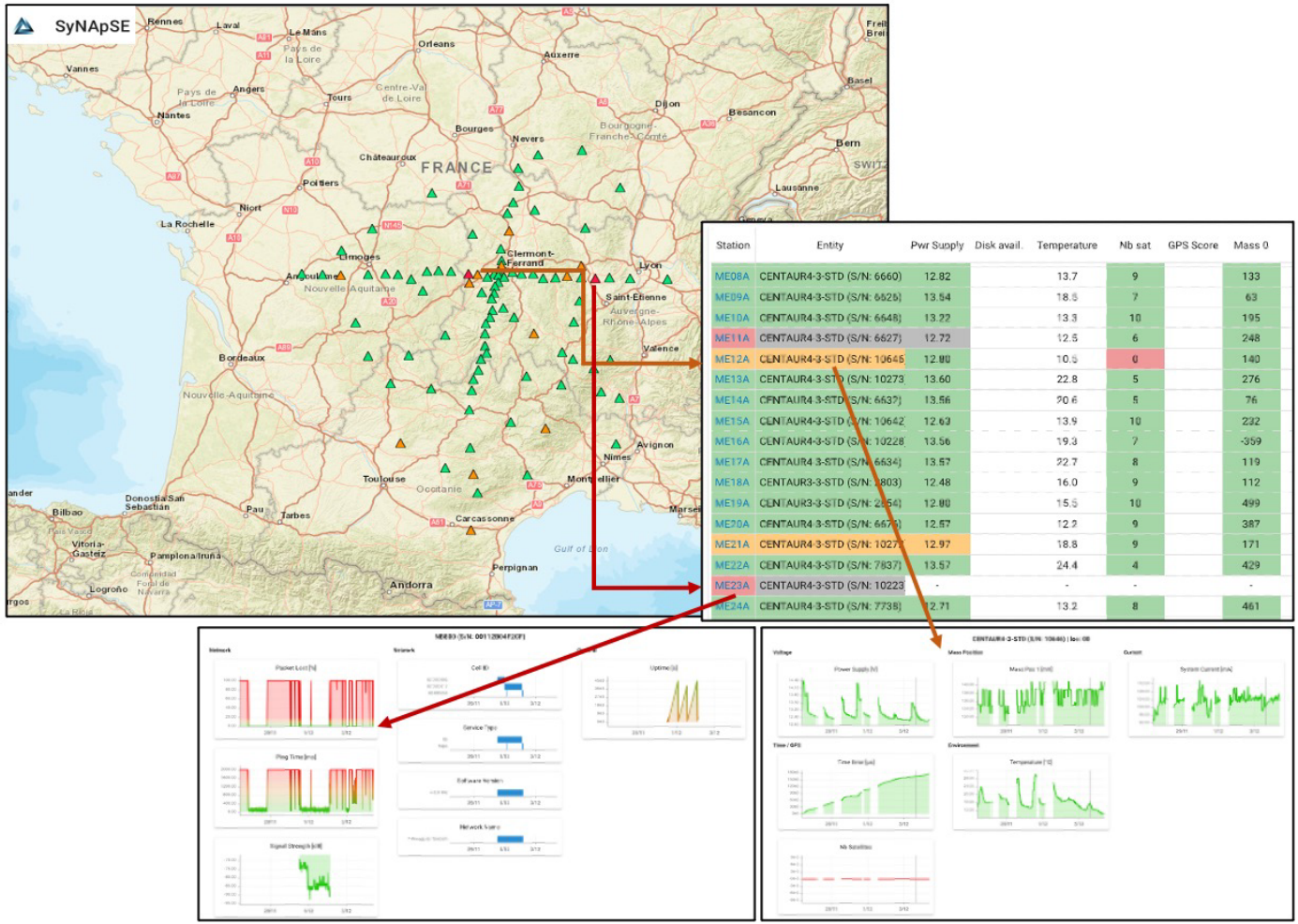

**Figure 6.** Illustration of SYNAPSE tool operation showing examples of failures at 2 stations. XF.ME12A has a GPS signal reception problem resulting in a data synchronization error. Synchronization was initially configured using the NTP protocol before the faulty antenna was replaced. XF.ME23A shows a lack of data transmission, therefore a gap in the real-time data stream. A phone call was first made to the site owner to ensure that no visible damage had occurred (e.g. antenna cable cut), followed by a call to the mobile phone operator. A technical incident involving the operator's relay antenna is in progress.

These daily control procedures ensure optimum operation of the temporary networks. The data availability plot calculated for XP shows a data completeness close to 100% for 32 of the 35 stations over an operating period ranging from station installation in 2023 to 2024-12-01. The larger data gaps in stations XP.FR18A, XP.FR19A and XP.FR26A are due to power cuts, a cut of the LTE and GNSS antenna cables (vandalism), and a sensor default, respectively (Fig. 7). The data availability plot of XF network is calculated for June to January 2025. It shows a good data completion, with scores at or very close to 100% for the 62 stations installed to-date. The few gaps at XF.ME03A, XF.ME11A, XF.ME12A, XF.MN14A, XF.MN17A are due to digitizer defaults, most certainly linked to GPS signal loss (Fig. 8).

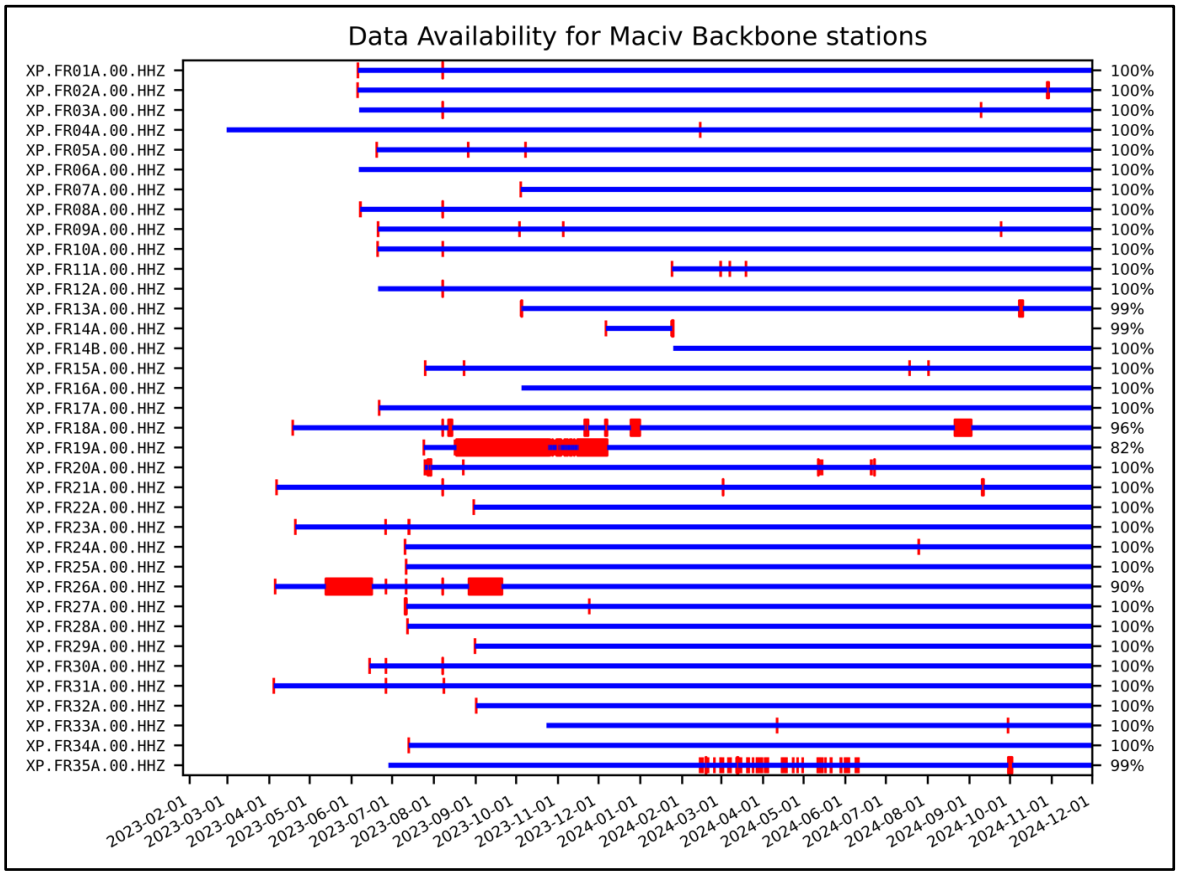

**Figure 7.** Data availability as a function of time for the XP network, from installation day to end of November 2024. Data gaps are shown in red. Data completion scores (in %) appear in the right-hand column. Gaps at the beginning of the acquisition period are due to installation.

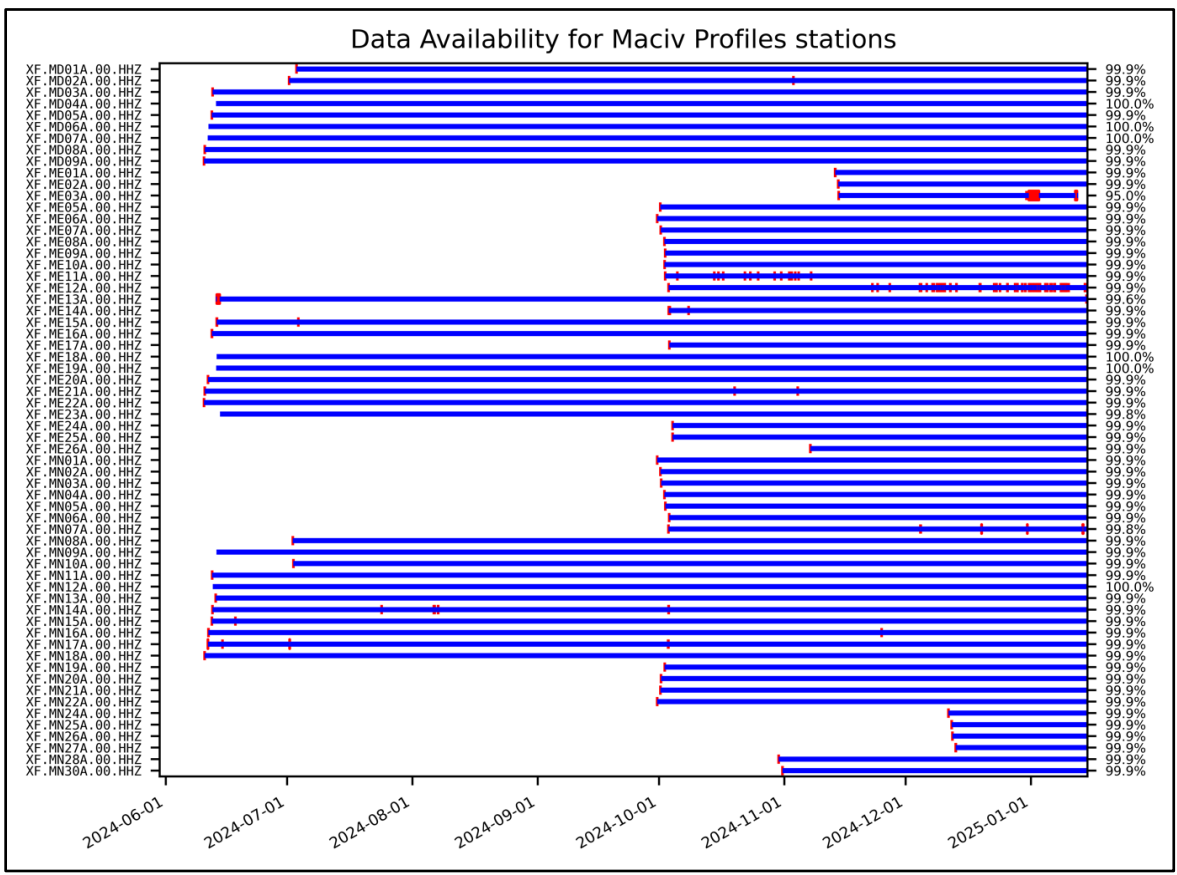

**Figure 8.** Data availability as a function of time for 62 intermediate-band stations of the XF network, from the installation day to the 15th of January 2025. Same legend as Fig. 7.

Another data retrievability test was carried out using the eida-data-monitoring procedure developed by J. Stampa at Kiel University (Stampa, 2023; Kolínský, Stampa et al., 2025). This procedure consists in retrieving random blocks of data over a period of time by querying the data center webservices. We applied this procedure to all temporary stations, retrieving 20 days of data in a period of 2 months by querying the Epos France SI-S webservice (French EIDA node). The associated data metadata are also retrieved, and the data is deconvolved from the instrumental response. This test was repeated three times over 2-month periods since the beginning of 2024: January - February 2024, June - July 2024, October - November 2024. The test not only verifies the presence of data at the data center, but also the conformity of the associated metadata. These successive tests show that data retrievability on the French EIDA node for the XP and XF networks is very good (close to 100%) throughout the tested period (Fig. 9). The related metadata are also present.

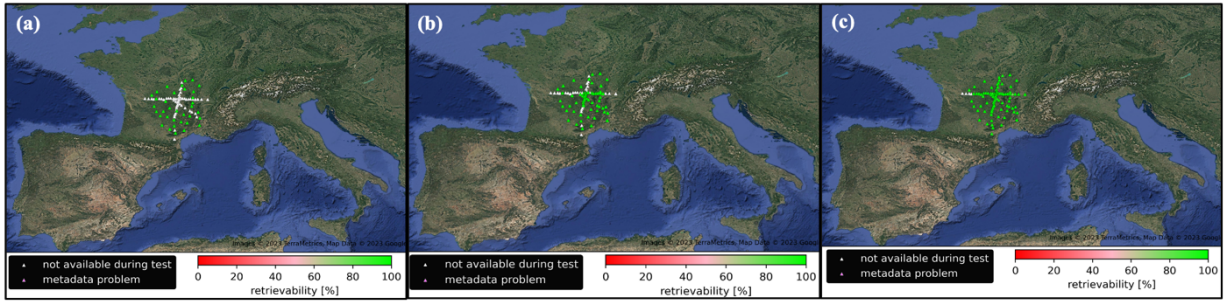

**Figure 9.** Data retrievability test performed with the eida-data-monitoring tool over 3 periods of time: (a) January – February 2024, (b) June – July 2024, (c) October – November 2024. Green triangles show results of data retrievability above 80%, white triangles show stations which were not yet installed when the test was conducted.

## 3.6  Data quality check

### 3.6.1 Noise analysis and data validation using Power Spectral Density (PSD)

Once a station is installed, we initiate a data quality check to validate the installation, the data and metadata. We first calculate Power Spectral Density (PSD) curves over a period of 1-2 weeks following deployment, using the clb-noise-analysis software (Vergne & Bonaimé, https://gitlab.com/resif/clb-noise-analysis).

By generating probability density functions of the PSD (PPSD) and spectrograms, this software provides a comprehensive view of noise levels and their temporal variations (Fig. 10). The analysis of these curves allows to identify noise sources or instrument anomalies, and to check the noise level with respect to reference noise models, such as the New Low Noise Model (NLNM) and the New High Noise Model (NHNM) (Peterson, 1993).

This fast computation of PSDs right after installation led us to move one station, XP.FR14A, where we observed a high noise level at low frequencies (<0.1 Hz), particularly on the horizontal components. Initially installed in a small outdoor shed with poor thermal insulation, the station was very sensitive to wind and temperature changes. It was relocated in a new indoor location with a more stable environment, resulting in a substantial improvement in noise levels. The new station code is XP.FR14B, in line with the AlpArray rule that the last character is replaced by B when the station is moved by more than 10m. The recordings of XP.FR14B now meet the expected quality standards for a broadband station.

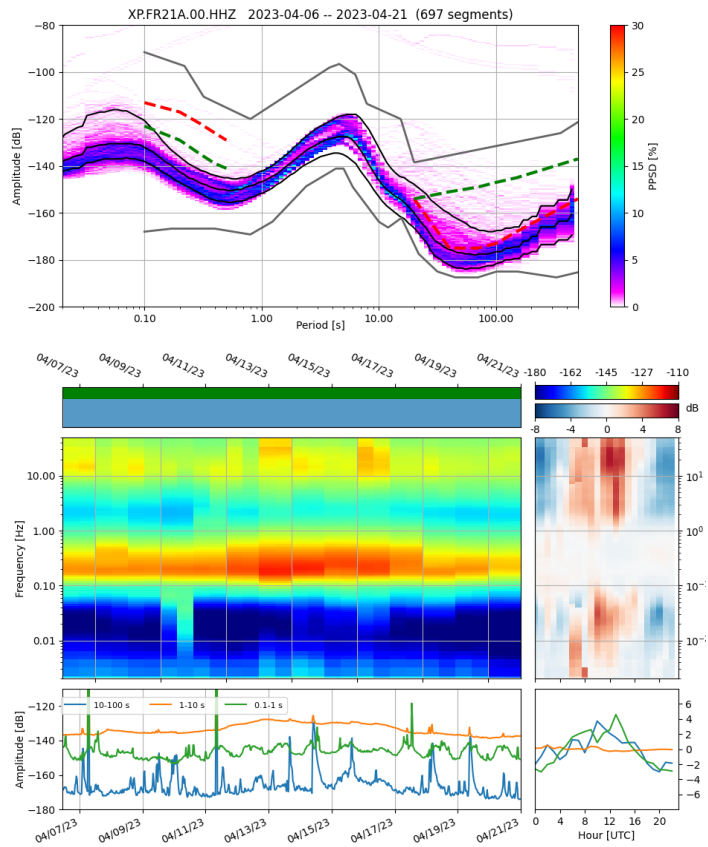

**Figure 10.** Example of result of the clb-noise-analysis software. Top: PPSD of XP.FR21A for 15 days after installation (2023-04-06 to 2023-04-21). Left col.: spectrogram, and amplitude of noise in 3 frequency bands. Right col.: daily mean of spectrogram, and mean of amplitude of noise in 3 frequency bands.

The PPSDs were systematically computed for the XP network over a one-year period, from September 2023 (or from their installation date if later) to August 2024. These computations were carried out using the clb-noise-analysis software, which also provides the median PSD curve over the time period. The median PSD curves were then compared with those from 11 permanent stations in the FMC to assess the overall quality and performance of the XP network (Fig. 11). The median PSDs of the vertical components (HHZ) generally show similar noise levels and spectral characteristics as permanent stations, 20–30 dB below the NHNM over the entire period range. However, the horizontal components of temporary stations consistently exhibit higher noise levels than permanent stations at long periods > 20 s. This difference is due to increased exposure of temporary stations to meteorological factors because they are installed on the ground while permanent stations are installed in boreholes (Vergne et al. 2019). Indeed, a single permanent station, FR.BANN, which sensor is buried a few cm in a PVC tube and not in a borehole, has comparable noise levels to those of XP temporary stations (Fig. 11 - dashed blue line). A few XP stations even exhibit lower noise levels on the horizontal components than FR.BANN.

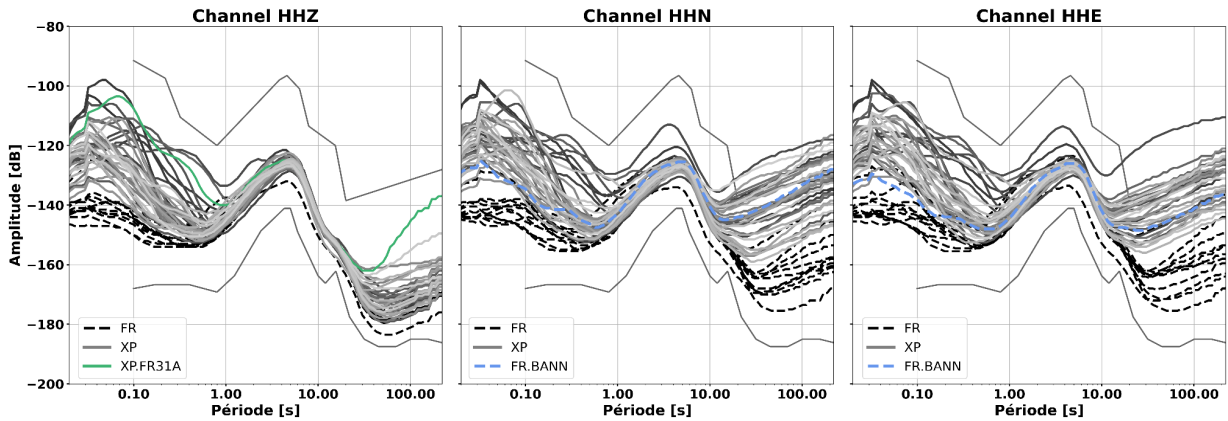

**Figure 11.** Median PSD curves for broadband stations XP (35 stations - solid lines) and FR (12 stations - dashed lines) from September 2023 (or from the installation) to the end of August 2024. Each line corresponds to a single station.

We however identified an anomalous high noise level at periods >100 s on the Z component of station XP.FR31A (green solid line in Fig. 11 left), as compared to other XP stations (grey solid lines). The analysis of the spectrogram at XP.FR31A over a 15-day period in summer highlights a recurring daily noise pattern, clearly visible in Figure 12b. In contrast, this noise is attenuated in winter, as shown in Figure 12a. These day/night variations show that the station, which is located outdoor, is highly sensitive to temperature cycles and weather conditions due to poor thermal insulation of the sensor. To mitigate these disturbances and improve data quality, insulation work is planned in the coming months.

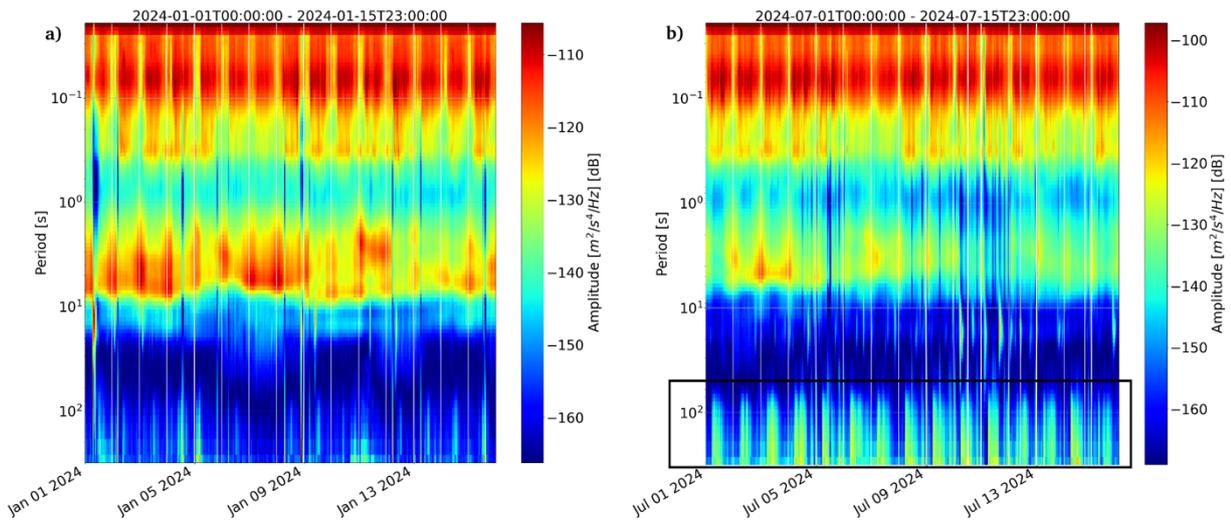

**Figure 12.** Spectrograms of the Z comp. of XP.FR31A: (a) from 2024-01-01 to 2024-01-15; (b) from 2024-07-01 to 2024-07-15. The black rectangle in (b) highlights the daylight noise at low frequencies.

Overall, despite expected higher noise levels on the horizontal components, the XP network has reliable performance across a wide range of frequencies, with vertical components performing particularly well. The network design and installation conditions contribute to its robustness and its ability to provide high-quality waveforms.

### 3.6.2 Orientations of horizontal components

An additional step in our data quality control process is the verification of the orientations of horizontal component. A preliminary check was conducted by analyzing receiver function (RF) waveforms for the XP, XF, and FR networks for a few teleseismic events (see section 4.4). We identified orientation errors of the horizontal components in three stations: XP.FR15A, XF.MN04, and XF.MD05A. The components initially labeled HHN-HHE were updated to HH1-HH2, and the StationXML files were revised to indicate the estimated orientations (errors of 90° or 180°).

We will later measure more precisely the actual orientations of the horizontal components by using the AutoStatsQ code developed by Petersen et al. (2019), which can estimate the orientations of horizontal components through Rayleigh wave polarization analysis.

# 4. Data examples

## 4.1 Records of major teleseismic and main regional earthquakes

We document the data quality of MACIV temporary stations by showing record sections of three earthquakes, one at teleseismic distance and two at regional distances.

Figure 13 shows records of a M7.0 teleseismic event that occurred in northern California on Dec. 5, 2024. The vertical component records of the 94 MACIV stations installed at the time are shown in a record section in Fig. 13a. Figure 13b shows the epicentral location with respect to the network location, and Fig. 13c is a snapshot of Rayleigh waves, at 2417s after earthquake origin time, across the MACIV temporary networks and the permanent BB networks in France. This set of figures (13a, b and c) is a seismic report generated for each major earthquake, sent to project participants for monitoring purposes, and integrated into the project website for dissemination (https://maciv.osug.fr/-Seismes-remarquables- , in French). Figure 14 shows records of a M4.2 regional earthquake that occurred in the Italian Western Alps on Dec. 9, 2024. The event was recorded by 94 temporary MACIV stations.

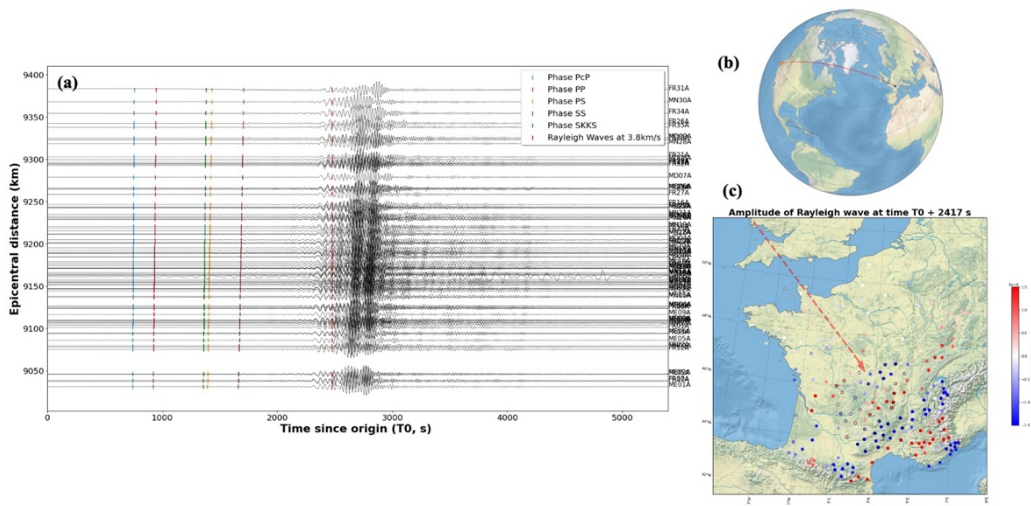

**Figure 13.** Vertical component records of the M7.0 teleseismic event of 2024-12-05 in Northern California by the MACIV temporary networks (XP and XF). (a) Record section with 0.008-0.05 Hz bandpass filter. The theoretical arrival times of the main body wave phases are displayed as color line segments, as well as the arrival time of the Rayleigh wave with 3.8 km/s velocity used in (c) (in red). (b) Location map of the earthquake and great circle to the MACIV arrays. (c) Snapshot of Rayleigh wave amplitudes 2417 s after the earthquake origin time (red line segments in (a)), at all permanent broadband stations in France (FR network) and temporary stations of the MACIV backbone (XP, stations shown by circles outlined in black). Amplitudes in μm/s.

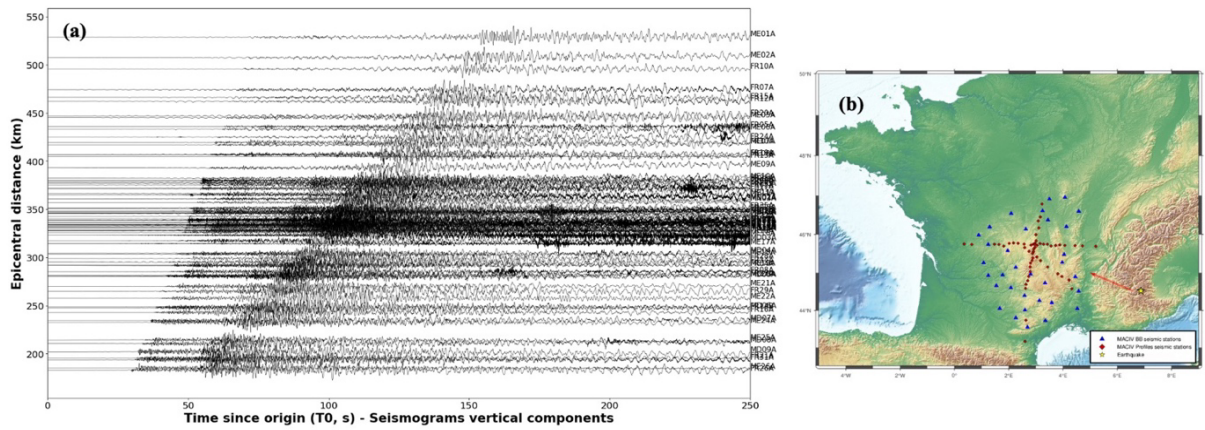

**Figure 14.** Vertical component records of the M4.2 regional earthquake of 2024-12-09 in Northern Italy, by the XP and XF stations. (a) Record section, with 0.3-10 Hz bandpass filter. (b) Location map of epicenter and MACIV stations.

## 4.2 Records of the Greenland seiche event of 2023-09-16

Figure 15 shows records by XP stations of the first half-day of the surface wavetrain generated by the seiche event of 2023-09-16 in Eastern Greenland. According to Svennevig et al. (2024), a rock-ice avalanche in a fjord generated a tsunami that stabilized into a seiche event of high amplitude and long duration, with equivalent moment magnitude (Mw) of 4.3. The seiche event radiated a monochromatic signal of 92 s period that lasted for 9 days with low amplitude decay. It was clearly recorded by the XP temporary network.

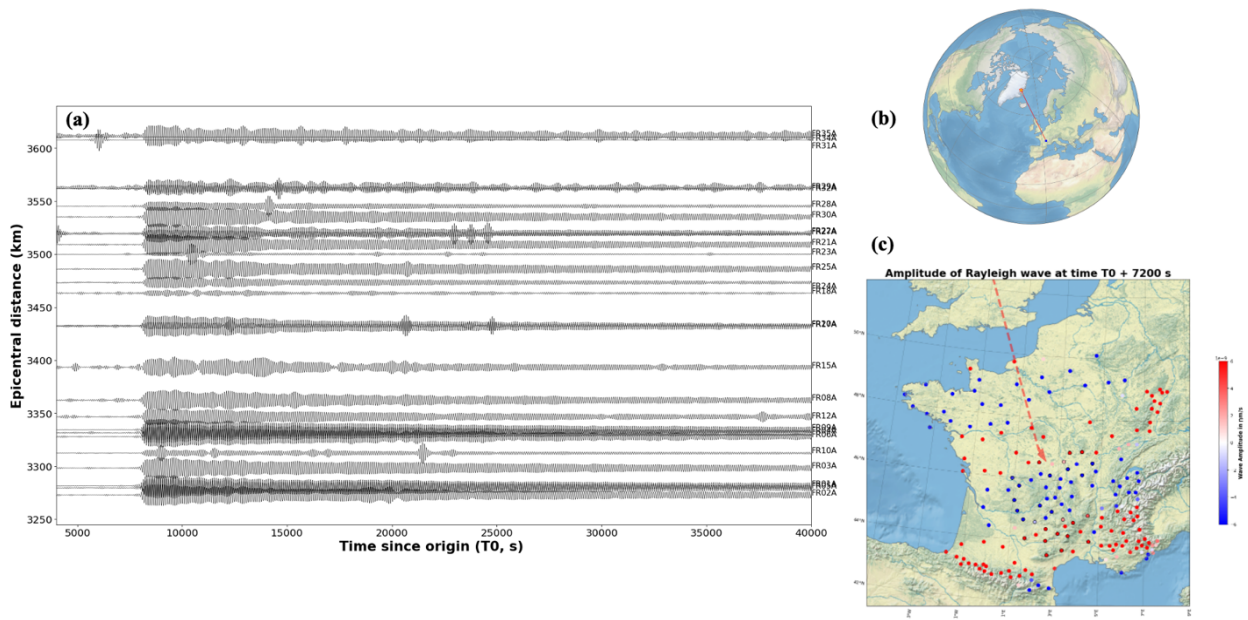

**Figure 15.** Vertical component records by the XP network of the Greenland seiche event of in 2023-09-16, 12:35:00 UTC. (a) Section plot showing the 13 first hours of the event (bandpass filtered signals between 0.01 and 0.0125 Hz). (b) Location map of the event and great circle path to the MACIV arrays. (c) Snapshot of Rayleigh wave amplitudes 7200s after the event origin time at all permanent broadband stations in France (FR network) and temporary stations of the MACIV backbone (XP, stations shown by circles outlined in black). Amplitudes in nm/s.

## 4.3 Earthquake location with SeiscomP: preliminary results

In June 2024, an earthquake (EQ) location system based on the SeisComp software package (Helmholtz Centre Potsdam GFZ German Research Centre for Geosciences and Gempa GmbH, 2008) was installed on a dedicated virtual machine at ISTerre (Fig. 5). Systematic EQ location in the French Massif Central (FMC) is currently being done on a monthly basis, with a weekly target by early 2025, using records of all temporary stations of the MACIV project and 56 stations of French permanent networks.

857 earthquakes have been processed between June 2023 and December 31, 2024 (Fig. 16), which outline the main seismicity zones of the FMC (Beucler et al. 2021). The Combrailles area northwest of Clermont-Ferrand, on either side of the Sillon Houiller is the most active zone, both in terms of number of events and magnitude. The Mont Dore region is characterized by EQ swarms, as the Ambert graben where several earthquake swarms are observed including the one of May 16 to May 20, 2024 with 271 events (inset in Fig. 16). The Saint-Flour region includes both clustered and diffuse seismic activity, and the near-Guéret zone has persistent seismicity which makes the link with Armorican seismicity. These preliminary results open up very interesting research prospects.

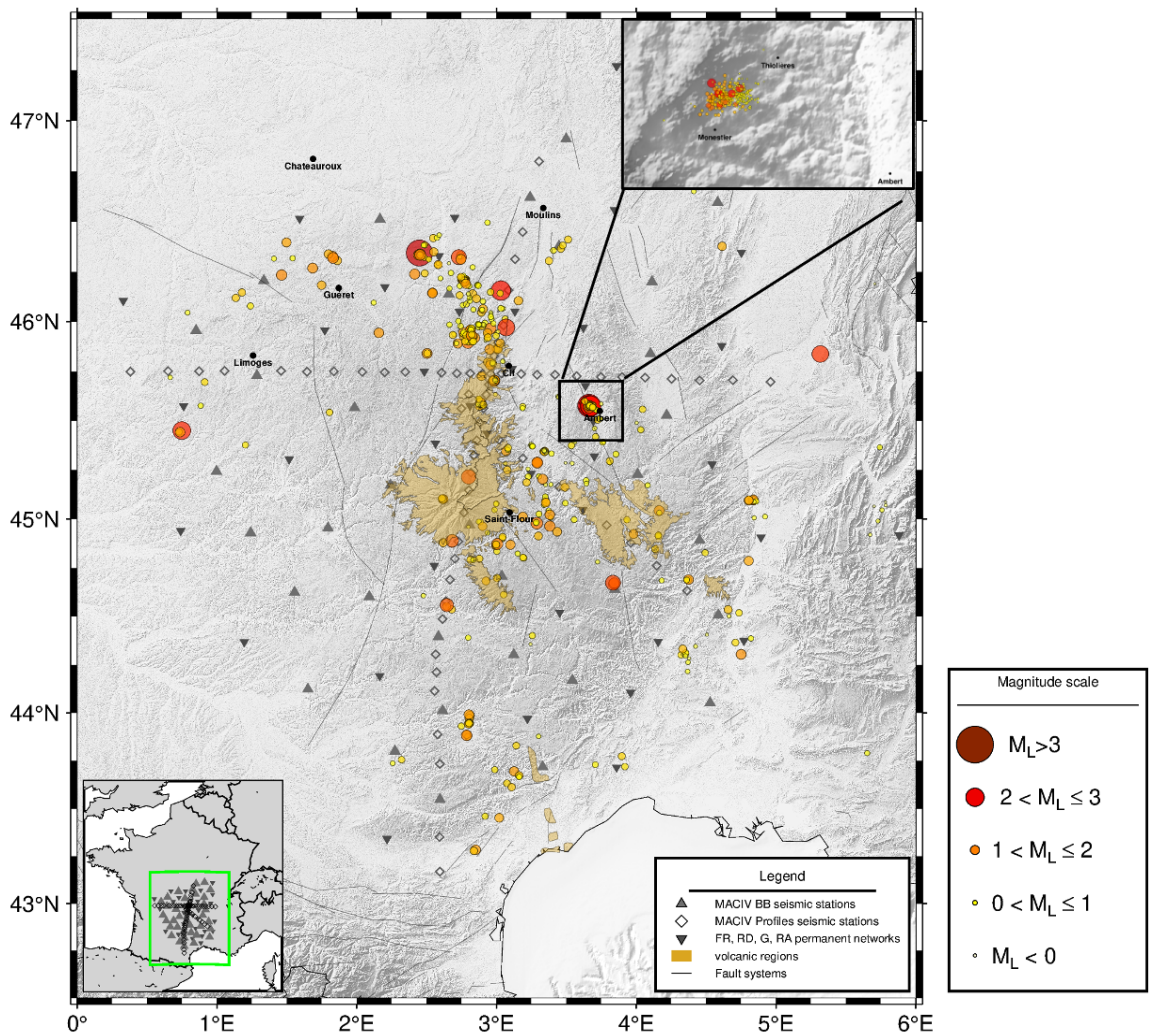

**Figure 16.** Seismicity map of the French Massif Central showing earthquakes recorded from June 1, 2023 to December 31, 2024 (circles with size depending on magnitude). Triangles and diamonds show the MACIV temporary stations, while upside-down triangles show permanent stations used by the monitoring system (networks code FR, RD, G, RA). The volcanic regions of the FMC are shown in yellow and the main fault systems by black lines. A zoom-in on the Ambert area highlights the 271 earthquakes of the May 2024 swarm.

## 4.4 Preliminary receiver function analysis

To quickly check the orientations of horizontal components, we computed receiver functions for the largest-magnitude teleseismic events. Figure 17 shows radial receiver functions computed for the M6.8 earthquake of 2024-11-10 in Cuba region, at stations located along the N-S linear profile that follows the N-S transect of XF temporary stations (black line in the map of the left hand side). The strong amplitude arrival at t~0 s is the direct P wave, marked P. The P-to-S converted at the Moho at ~30 km depth is the positive phase at

t~4s, marked Ps. The multiple of the Moho converted phase, marked Pps, is clearly visible at t=12-13 s.

Receiver functions have helped to detect misorientations of horizontal components at stations XP.FR15A, XF.MN04, and XF.MD05A, as explained in section 3.6.2.

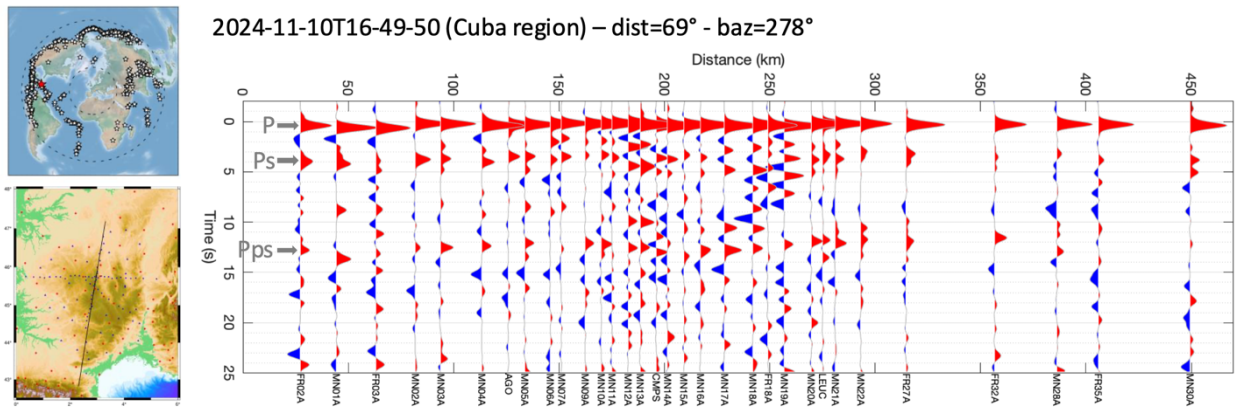

**Figure 17.** Radial receiver function profile along the N-S transect for the earthquake of 2024-11-10, M6.8 in Cuba region. Top left: event location (red star). Bottom left: station map; the black line shows the location of the profile. Right: radial receiver function waveforms computed for stations located along the N-S profile. Each waveform is plotted at the abscissa of the projection of the corresponding station on the profile. Distances are measured from the northern end of the line. Station codes shown at the bottom of the figure.

# 5. Conclusions

The MACIV project's temporary XP and XF networks have benefited from over 15 years of continuous improvement in broadband station installation procedures in France and neighboring countries by our teams. The first few months of operation have demonstrated the reliability of our new installation procedures, documented by a rather low noise level at long periods on the vertical components (-180 to -160 dB at T>20 s) and a higher noise on the horizontal components (-150 to -120 dB), which is generally the case with unburied broadband velocimeters. The value of GSM telemetry is high, not only for distributing waveforms in near-real time, but also for state-of-health monitoring of temporary stations using the same tools as for permanent stations. As compared to previous large-scale temporary experiments, the use of reliable installation procedures and GSM communication greatly reduces the number of maintenance visits, saving time, money and $CO_2$ emissions.

The final phase of the MACIV multi-scale seismic experiments is the deployment in autumn 2025 of dense (0.5 to 7 km spacing) large-N networks dedicated to imaging the shallow structure of the quiescent volcanic regions of the Chaîne des Puys and Cézallier-Monts Dore.

Data analysis of the broadband seismic networks has just begun at the time of writing. We are confident that new information will emerge soon on the lithospheric structure of the French Massif Central and on the potential magmatic activity of its volcanoes, with applications to deep geothermal energy.

**Data availability statement.** Data can be downloaded at https://10.15778/resif.xp2023 (public data) and https://doi:10.15778/resif.xf2024 (embargoed until July 2026).

**Acknowledgements.** The MACIV project is funded by Agence Nationale de la Recherche, France (project ANR-22-CE49-0019). All seismic instruments used in the MACIV project belong to the French pool of mobile instruments SISMOB (Epos-France). We thank the Epos-France EIDA node for efficient data archiving and distribution. We are grateful to the team responsible for the SYNAPSE tool for their rapid integration of MACIV stations. We are grateful to the many individuals and municipalities who helped us finding good sites for our temporary stations and who host them. We acknowledge support from our colleagues J. Battaglia, G. Boudoire, S. Duchêne, D. Laporte, and O. Vanderhaeghe who participate in the MACIV project.

**CORRESPONDING AUTHOR: Coralie AUBERT**,
ISTerre, Univ. Grenoble Alpes, France
e-mail: coralie.aubert@univ-grenoble-alpes.fr
+33 6 14 83 07 05